# The properties of ultrapure delafossite metals

A.P. Mackenzie

*Max Planck Institute for Chemical Physics of Solids, Nöthnitzer Strasse 40, 01187 Dresden, Germany*

and

*Scottish Universitites Physics Alliance, School of Physics & Astronomy, University of St Andrews, North Haugh, St Andrews KY16 9SS, Scotland*

**Abstract**

Although they were first synthesized in chemistry laboratories nearly fifty years ago, the physical properties of the metals $PdCoO_2$, $PtCoO_2$ and $PdCrO_2$ have only more recently been studied in detail. The delafossite structure contains triangular co-ordinated atomic layers, and electrical transport in the delafossite metals is strongly two-dimensional. Their most notable feature is their in-plane conductivity, which is amazingly high for oxide metals. At room temperature, the conductivity of non-magnetic $PdCoO_2$ and $PtCoO_2$ is higher per carrier than those of any alkali metal and even the most conductive elements, copper and silver. At low temperatures the best crystals have resistivities of a few $n\Omega cm$, corresponding to mean free paths of tens of microns. $PdCrO_2$ is a frustrated antiferromagnetic metal, with magnetic scattering contributing to the resistivity at high temperatures and small gaps opening in the Fermi surface below the Néel temperature. There is good evidence that electronic correlations are weak in the Pd/Pt layers but strong in the Co/Cr layers; indeed the Cr layer in $PdCrO_2$ is thought to be a Mott insulator. The delafossite metals therefore act like natural heterostructures between strongly correlated and nearly free electron sub-systems. Combined with the extremely high conductivity, they provide many opportunities to study electrical transport and other physical properties in new regimes. The purpose of this review is to describe current knowledge of these fascinating materials and set the scene for what is likely to be a considerable amount of future research.

# 1. Motivation and historical overview

*1.1 Context and motivation*

Condensed matter physics is a broad field covering both fundamental and applied science. For me, its excitement is the opportunity that it offers to test the beautiful underlying principles of many-body physics by studying collective assemblies of electrons. Thanks to advances in synthetic chemistry and measurement techniques, there has been a huge increase in the number of accessible materials and experiments. This quiet revolution in experimental capability has gone hand in hand with a series of profound theoretical developments, in which fascinating links have been uncovered between the behavior of electrons in materials and that of other forms of matter. It feels as if a slightly worrying, fifty-year long tendency for the sub-fields of science to separate from one another is beginning to reverse, and I believe that there is real cause for optimism for the coming decades.

The relative ease with which materials can now by synthesized and studied is both a blessing and a curse. The problem is that real materials are often highly complex, with very high levels of disorder that cannot be satisfactorily treated theoretically. This disorder is therefore like a fog obscuring the true underlying physics. There is therefore a premium on identifying simple model systems with long charge carrier mean free paths in which fundamental principles can be studied or even established. This explains the impact that such systems, for example graphene, have when they are discovered.

The goal of this short review is to assess the prospects for layered triangular lattice metals, specifically those adopting the delafossite structure, to become benchmark systems for condensed matter physics. Many reviews are written as fields reach maturity, to sum up the collective achievements of large bodies of work. This article has a different goal. My aim is summarize what is currently known in what is in some senses still a very young sub-field, with a view to stimulating future work.

*1.2 Brief historical review*

The assertion that the study of delafossite metals is still in its infancy runs into an immediate problem with the first citations that any review of the field must give. The materials that will be covered in the bulk of this article were first reported in three seminal papers from a group at the DuPont Experimental Station, published in 1971 [1–3]. Shannon, Prewitt and Rogers reported the synthesis, basic crystal chemistry and physical properties of no fewer than nineteen oxides with general formula $ABO_2$ (A = Pt, Pd, Ag or Cu, B = Cr, Co, Fe, Rh, Al, Ga, Sc, In or Tl) and the structure pictured in Fig. 1. Among these, nine were synthesized for the first time, and nine were grown in single crystal form. [1]

The DuPont group research immediately illustrated one of the key features of the delafossite structure: in common with other classic structures of layered oxides, it is very accommodating, and has the flexibility to host many different elemental combinations. Indeed, other combinations have since been synthesized; a prominent example will be touched on later in this paper. Other crucial facts illustrated by the foundational work of Shannon and colleagues are that: a) most of the materials are non-metallic; b) most are close to stoichiometric, though some cation solid solutions exist; c) the Pd and Pt –based materials have Pd and Pt in an unusual 1+ formal valence state attributed to linear co-ordination with oxygen atoms directly above and below them in the delafossite structure. These Pd and Pt-based materials dominate the metallic delafossites; the Ag and Cu-based materials are usually semiconductors.

---

[1] The delafossites can actually exist in two structural polymorphs, termed 3*R* and 2*H*. The Pd- and Pt-based metals discussed here all crystallize in the more common 3*R* polymorph, whose structure is the one sketched in Fig. 1. For a convenient review of the crystal chemistry of the overall series I refer the reader to [4].

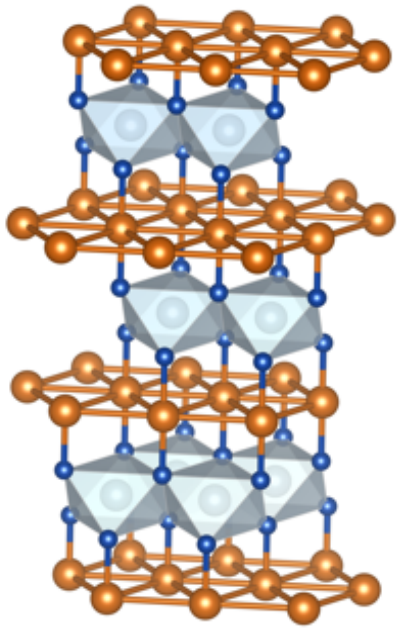

**Fig. 1** (colour online) The delafossite crystal structure for materials with the general formula $ABO_2$. The A site cations are the large spheres, the oxygens are the small spheres and the B site cation sits in the middle of the oxygen atoms. The linear co-ordination of the A site cation between two oxygens is clearly seen. Figure courtesy of V. Sunko.

Why, then, can the study of delafossite metals still be regarded as a 'young' field? The answer is that although research on the semiconducting materials was taken up by many groups, almost no work was done on the metals. This is surprising because the original DuPont papers highlighted the extremely high room temperature in-plane conductivity of $PtCoO_2$, quoting a value of 3 µΩcm and noting that it was comparable to that of elemental copper. Shannon and co-workers also stated without comment the value of 2 µΩcm for the room temperature resistivity of $PdCoO_2$. This seems to have been widely overlooked by physics groups, with materials like $IrO_2$, $ReO_3$ and $SrMoO_3$ with respective room temperature resistivities of 24 µΩcm, 8 µΩcm and 5 µΩcm becoming known as benchmark highly conducting oxide metals [5–8]. One can only speculate about the reason for this, but several factors might have come into play. The data reported in 1971 were only at room temperature, though ref. 3 contained the statement that resistivity ratios between 10 and 100 were seen across the range of crystals studied, and the crystals from which the resistivity results were obtained were small and not easy to grow. Also, the large resistance ratio in excess of 100 meant that rather high voltage resolution would be required to even measure the temperature dependent resistivity to low temperatures with confidence.

Whether for the above reasons or just accidentally, metallic delafossites were essentially ignored for over two decades. Then, in the mid 1990s, Tanaka and co-workers at the University of Tokyo reported the growth of $PdCoO_2$ and $PtCoO_2$ crystals [9,10] and the first measurement of the temperature-dependent resistivity of $PdCoO_2$. The in-plane resistance ratio of those crystals was a modest 10, and the magnetic susceptibility also gave evidence for a substantial impurity level [10], but the group confirmed that the conductivity was strongly anisotropic, with in-plane conductivity several hundred times larger than that perpendicular to the planes, and also progressed understanding of the materials in a series of spectroscopic papers that will be reviewed later in this article.

The work of Tanaka and colleagues attracted the attention of people with interest in layered metallic oxides, but the next substantial advance came when Takatsu, Maeno and co-workers at Kyoto University began work on delafossite metals a decade later. Motivated by their discovery of superconductivity in the nearly free electron oxide $Ag_5Pb_2O_6$ [11], they first grew new crystals of $PdCoO_2$ with substantially increased in-plane resistance ratios of over 400 [12], and then grew the first ever single crystals of the related compound $PdCrO_2$, with in-plane resistance ratios of over 200 [13]. No superconductivity was found, but many other significant phenomena have been observed in the two materials, and the knowledge that such high crystal quality was attainable stimulated the interest of many other groups. In that sense, the Kyoto group's work kick-started contemporary research on delafossite metals, and led to the growing body of literature that will be reviewed in this paper.

## 2. Why delafossite metals are notable

In order to appreciate why the delafossite metals are so notable, it is instructive to continue the discussion of their electrical conductivity. To set this in context, the room temperature resistivities of a number of the most highly conductive materials known are depicted graphically in Fig. 2. The left panel shows the two well-known oxide conductors $IrO_2$ and $ReO_3$, compared with the alkali metals, elemental Pt and Pd and a number of the highest conductivity metals known clustered at the bottom of the figure. On the right panel, the lower cluster of materials is expanded, and it is seen that $PdCoO_2$ [14] and $PtCoO_2$ [15] fall among the best metallic conductors such Al, Au, Cu and Ag. However, the charge carrier density of the two layered delafossite metals is approximately a factor of three lower than that of the three-dimensional elements. The room temperature mean free paths of $PtCoO_2$ and $PdCoO_2$, approximately 700 Å and 600 Å respectively, are therefore the longest of any known large carrier density metal[2].

The situation is arguably even more surprising when one considers the low temperature resistivity. The lowest published value is 0.0075 μΩcm for $PdCoO_2$, corresponding to a mean free path $\ell$ of ∼ 20 μm [14], extremely large for any metallic compound and a world record for an oxide. At low temperatures, the collisions that efficiently relax electron momentum and lead to resistivity take place only from static disorder, usually assumed to be point defects, e.g. vacancies or impurity atoms in the conducting planes. To put $\ell$ in $PdCoO_2$ in context, consider the following comparisons with other materials. Careful annealing and material refinement can give mean free paths of mm or more in metallic elements (though they need to be treated very carefully to avoid introduction of dislocations). As soon as compounds or quasi-two dimensional materials are considered, however, the situation is very different. The cuprate superconductors in which the de Haas-van Alphen effect has been seen have $\ell$ of order a thousand times lower than that of PdCoO$_2$ [16]. Years of careful material refinement have led to $\ell \sim 1.5$ μm in the highly purity-sensitive superconductor Sr$_2$RuO$_4$ [17] and the ZnO-(Zn,Mg)O heterostructures that have recently shown the Fractional Quantum Hall Effect [18]. Only very recently has a mean free path in graphene as high as 20 μm been achieved [19,20], and achieving it in semiconductor heterostructures took decades of painstaking research. It is remarkable that such a value is observed in crystals of PdCoO$_2$ grown in hot crucibles using modified flux methods and without further refinement or purification. Using the known areal density of Pd atoms (1.45 x 10$^{15}$ cm$^{-2}$) and a simple model assumption of one strongly scattering defect in an area of $\pi\ell^2$ implies one defect per 2 x 10$^{10}$ lattice sites, a truly remarkable (and chemically implausible) defect density.

---

[2] The qualification of large density is made because lightly doped graphene has a weakly temperature-dependent resistivity and a very long mean free path of order 1 μm even at room temperature. Another prominent example of an element with long mean free paths is the semi-metal bismuth. Its room temperature carrier concentration is at least $10^4$ lower than the metals included in this comparison so although its room temperature resistivity is over 100 μΩcm, its mean free path is longer than that of the high carrier density metals included in the comparison given here.

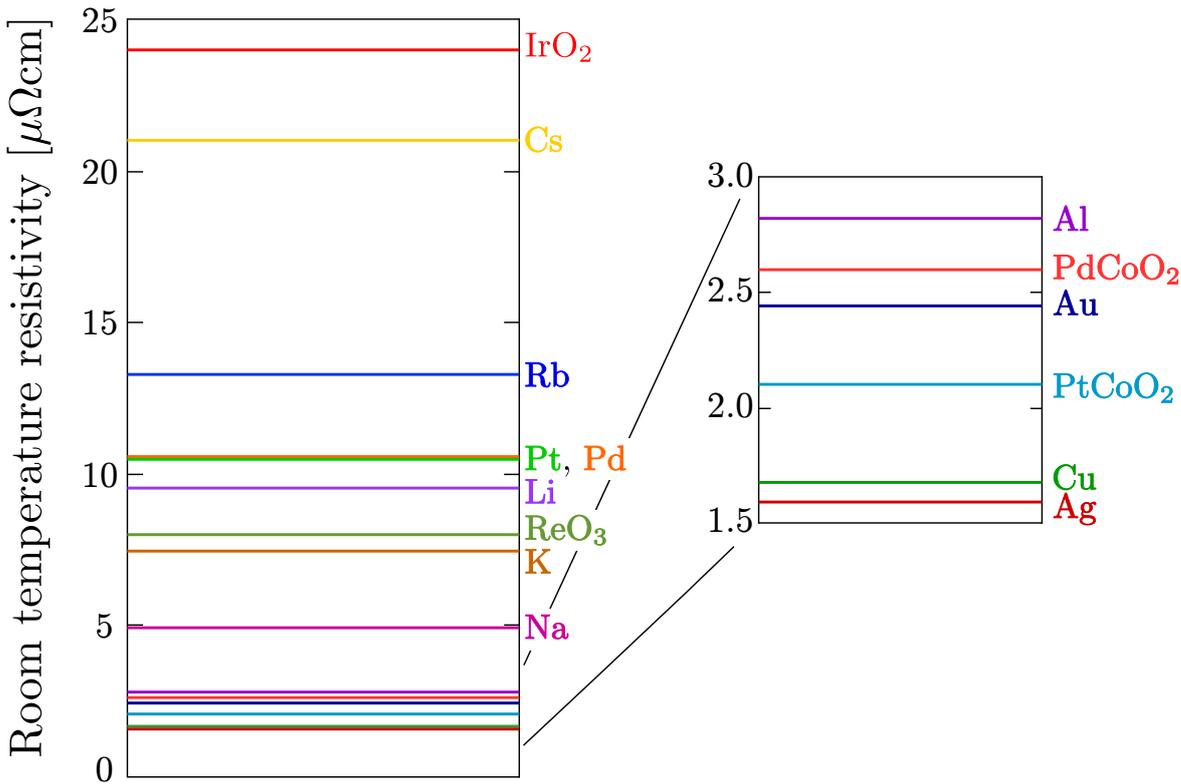

**Fig. 2** (colour online) Graphical depiction of the room temperature resistivities of the most highly conducting metals known. Figure courtesy of V. Sunko.

Although the reasons for the above observations are not yet fully understood, the delafossite metals therefore easily satisfy the long mean free path criterion for benchmark materials for modern condensed matter physics. They also satisfy the criterion of simplicity. As illustrated in Fig. 3 for $PdCoO_2$, a single highly two-dimensional band crosses the Fermi level, producing a cylindrical Fermi surface of nearly hexagonal cross-section. This basic Fermi surface topography has been confirmed experimentally in $PdCoO_2$, $PtCoO_2$ and (above 40 K) in $PdCrO_2$.

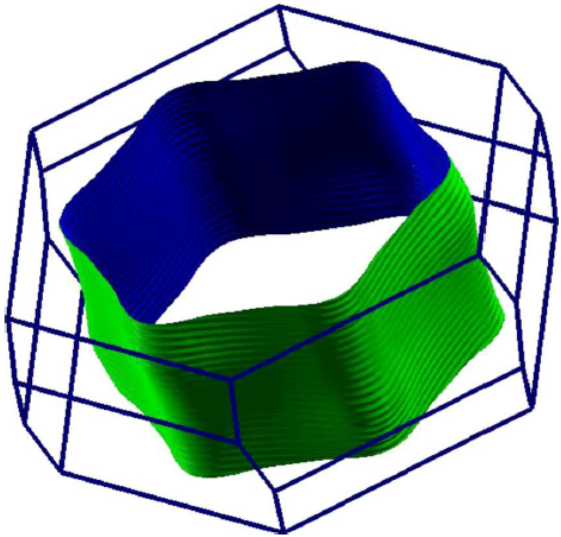

**Fig. 3** (colour online) The calculated Fermi surface of $PdCoO_2$, from Ong *et al.* [21]. It is nearly cylindrical, with an approximately hexagonal cross-section. Similar Fermi surfaces are calculated for the other known Pd- and Pt-based delafossite metals.

In addition to the purity and simplicity highlighted in Figs. 2 and 3, the chemical flexibility of the delafossite series also leads to an intriguing interplay between correlated and nearly free electron physics in some of the materials. Crudely, the delafossite metals can be thought of as naturally forming heterostructures between insulating, strongly correlated transition metal oxide layers and triangular co-ordinated Pd or Pt layers containing highly mobile electrons. For example, the Fermi velocity of $PtCoO_2$ is 8.9 x $10^5$, ~ 80% of the free electron value [15]. The insulating $CoO_2$ layers in both $PtCoO_2$ and $PdCoO_2$ are non-magnetic, but in $PtCrO_2$ the $CrO_2$ layers carry substantial local moments. Local spins in a triangular lattice are unable to adopt a conventional antiferromagnetic configuration because it is impossible to reverse the spins simultaneously in all three directions. The antiferromagnetism is therefore said to be 'frustrated'. A common way to relieve this frustration is the formation of a commensurate state in which the spins rotate by 120° between adjacent sites, forming an ordered magnetic state but at a temperature much lower . The Cr spins in $PdCrO_2$ enter an ordered state of this general class just below 40 K, offering a rare opportunity to study the coupling between nearly free electrons and a frustrated antiferromagnet. $PdRhO_2$ is also known to be a metal, and assumed to be non-magnetic. Although no reports of single crystals exist so far, it has been predicted to have a significantly lower conduction electron bandwidth than its Co-based relatives, primarily because of its larger lattice parameter [22], adding to the diversity of the physical properties of the series. In the longer term the overall delafossite series, including the many insulating, magnetic and semiconducting members, looks to be attractive for artificial layer-by-layer growth using modern oxide thin film techniques.

## 3. Organisation of this review

In order to describe the physics behind the properties summarized above, this paper will proceed with a general discussion of the crystal and electronic structures of the delafossite metals, from both band structure and localized perspectives, before the known properties of $PdCoO_2$, $PtCoO_2$, $PdRhO_2$ and $PdCrO_2$ are reviewed in more detail in three further sections. $PtCoO_2$ and $PdRhO_2$ will be covered jointly because of similarities in the early studies of the two compounds and the fact that relatively few reports have appeared to date about either. The article will close with arguably the most important section in a young field such as this – a discussion of the outstanding issues and opportunities for further work.

## 4. Crystal and electronic structures

*4.1 Crystal structure and optical phonon energies*

The structural parameters of $PdCoO_2$, $PtCoO_2$, $PdRhO_2$ and $PdCrO_2$ are given in Table 1. All have the structure of the mineral delafossite, $CuFeO_2$, namely the space group $R\bar{3}m$ $(D_{3d}^5)$. This is similar to the rocksalt structure of $NaCoO_2$ (a well-known superconductor when intercalated with water [23]) and the ionic conductor $LiCoO_2$, but with a different layer stacking sequence of the transition metal oxide and metal layers. In delafossite, there are three inequivalent planes, with the A site metal sitting directly above and below oxygen atoms from the neighbouring B site octahedra. Each metal sits in a flat plane with triangular co-ordination. Both the lattice parameters and the phonon frequencies are similar for all materials, though the 7% in-plane bond length difference between $PdRhO_2$ and $PdCoO_2$ is notable.

| Material | $a$ (Å) | $c$ (Å) | $A_{1g}$ phonon (cm$^{-1}$) | $E_g$ phonon (cm$^{-1}$) |
|---|---|---|---|---|
| $PdCoO_2$ | 2.83 | 17.743 | 712 | 520 |
| $PtCoO_2$ | 2.82 | 17.808 | 728 | 520 |
| $PdRhO_2$ | 3.02 | 18.083 | 664 | 523 |
| $PdCrO_2$ | 2.93 | 18.087 | 710 | 556 |

**Table 1.** Lattice constants of $PdCoO_2$ [12], $PtCoO_2$ [15], $PdRhO_2$ [1] and $PdCrO_2$ [1], along with Raman active phonon wavenumbers taken from [24] for $PdCoO_2$, $PtCoO_2$ and $PdRhO_2$, and [25] for $PdCrO_2$. Infrared active phonon wavenumbers have also been calculated and, for $PdCrO_2$, measured in the cited papers.

*4.2 Basic features of electronic structure*

In the language of ionic bonding (not strictly applicable to metallic delafossites but widely used nevertheless), it has been known since the original work in the 1970s that $ABO_2$ delafossites exist in the formal valences $A^{1+}B^{3+}O^{2-}{}_2$. The question of whether or not a delafossite will be metallic is then determined by the relative energetics of the bands of the relevant atoms in these valence states. If the B site cation is $Co^{3+}$, it adopts the non-magnetic low-spin configuration, making the cobaltates at least nominally amenable to DFT calculations. For $Ag^{1+}$ or $Cu^{1+}$, the Fermi level ($E_F$) sits in a natural gap between band multiplets of Ag or Cu, and the bands nearest $E_F$ are those associated with the $BO_2$ octahedra. For $Co^{3+}$, these are filled, and the stoichiometric materials are predicted within band theory to be insulators or semiconductors.

However, the energetics associated with $Pd^{1+}$ or $Pt^{1+}$ are very different. For these A site cations, a broad band, mainly Pd- or Pt- derived, crosses $E_F$, leading to the calculated Fermi surface shown in Fig. 3 for $PdCoO_2$. Within this zeroeth order approximation, the Pd- and Pt- derived delafossite metals are not oxide metals in the traditional sense, but are essentially triangular lattice layers of Pt and Pd metal, spaced by insulating transition metal-oxygen octahedra.

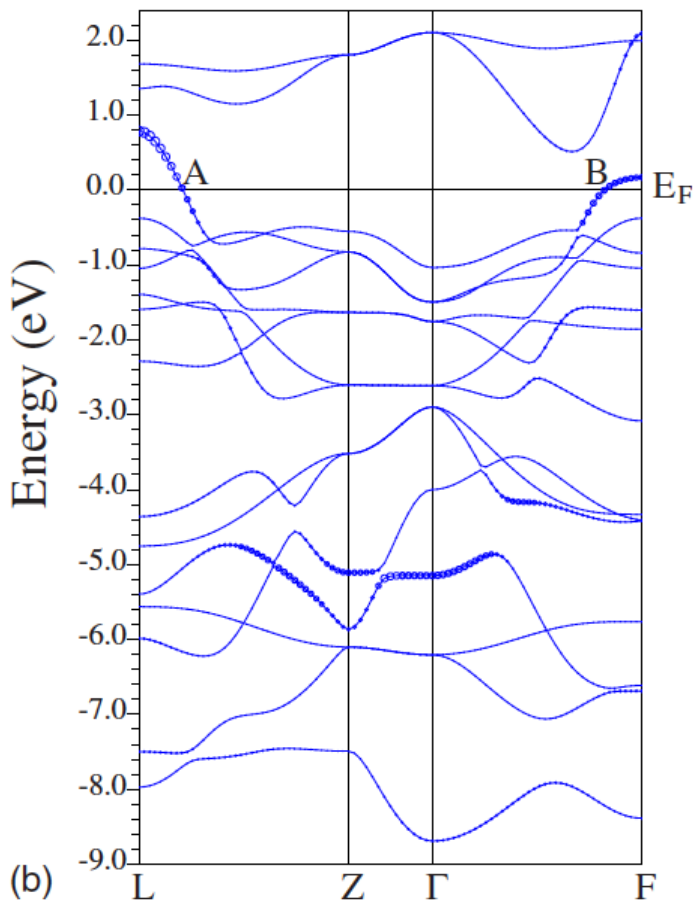

**Fig. 4** (colour online) The band structure of PdCoO$_2$, based on density functional calculations by Ong et al. [21]. The circles represent the calculated weight of a particular Pd d-orbital to the different bands, and points A and B represent two symmetry-distinct Fermi level crossings, but this level of detail is irrelevant to the main point, namely that a single band of substantially Pd character crosses the Fermi level.

The fact that DFT calculations predict filled Co-derived bands for PdCoO$_2$ means that the strength of correlation in those bands, which have strong 3*d* character, is not proven to be crucial in making them insulating. In PdCrO$_2$, however, the issue is fundamental. Simple electron counting in the absence of correlation would say that the Fermi level should now sit in narrow Cr-derived bands, something confirmed in an explicit calculation [26]. Ionic considerations, however, say that Cr$^{3+}$ should be in a spin-3/2 state due to on-site correlation, have a large local moment in real PdCrO$_2$ and that the CrO$_2$ layers should be correlated electron insulators. This is a point that I will return to in more detail in Section 7; for now I note that it is the viewpoint adopted, usually implicitly, in most of the literature, and that it seems to be supported by the majority of the available experimental evidence.

The fact that correlations dominate the physics of the B site electronic structure in the delafossites also seems to be consistent with the known properties of Ag and Cu-based delafossites with B-site cations other than Co$^{3+}$. In many of them, naïve electron counting allows the possibility of band insulators, but that same naïve counting predicts that AgFeO$_2$, AgCrO$_2$, CuCrO$_2$ and delafossite itself, CuFeO$_2$, should be metals. In contrast, they are magnetic insulators. An interesting exception to this trend is AgNiO$_2$, whose 2H structural polytype is observed to exhibit both metallic behavior and a series of charge ordering transitions. These, however, are also due to strong correlations associated with Ni, combined with the ability of Ni to adopt different valence states. Both the charge order and the metallicity are thought to come from the NiO$_2$ layers, associated with different patterns of Ni charge disproportionation [27]. Although some work on the metallic phases has been reported [28], research on metallic AgNiO$_2$ is still in its infancy, and since the cause of the conduction is different from that in all the other known delafossite metals, I do not discuss it in more detail in this paper. Hopefully research on it continues and it can be a topic for some future review.

Since the nearly hexagonal Fermi surface filling half the Brillouin zone is common to PdCoO$_2$ [21,29–31], PdRhO$_2$ [22] and PtCoO$_2$ [15,31] and is also observed at temperatures above 40 K in PdCrO$_2$ [32,33], it is useful to close this section with comments on how it arises in a tight-binding model. The first thing to note (Fig. 5a) is that it is not a feature of nearest-neighbour tight-binding on a 2D triangular lattice. A hexagon of the correct orientation does appear, but it is at ¾ filling, not ½ filling. A near-hexagon of the correct orientation at ½ filling appears if higher order hopping terms are taken into account. Figure 5b shows the constant energy contours for hopping to the next nearest neighbour along the bond direction (t$_3$=0.15 t$_1$); a similar shape of Fermi surface at half filling can also be obtained for hopping to the next-nearest neighbor, but with a negative sign (t$_2$/t$_1$<0).

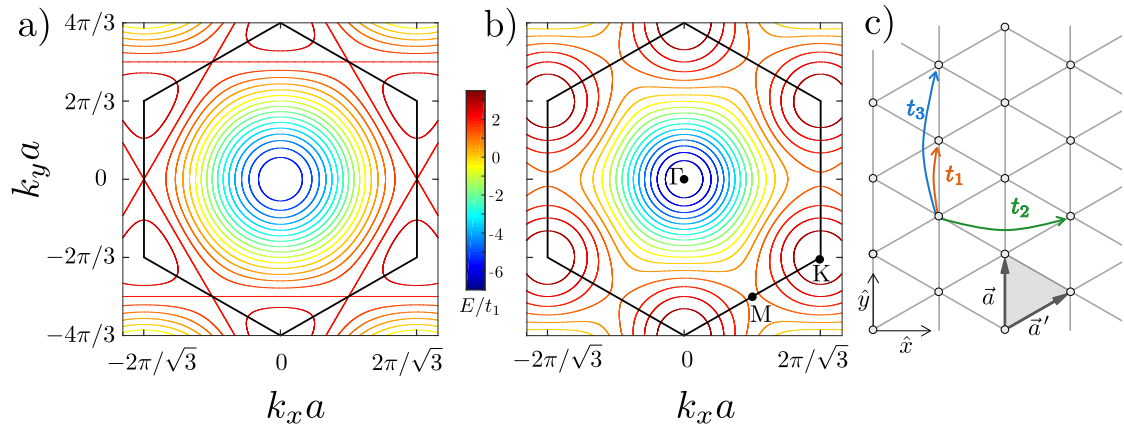

**Fig. 5** (colour online) a) Fermi surface contours for a nearest-neighbour tight binding model on a triangular lattice in two dimensions. A hexagon rotated by 30° from the Brillouin zone boundary appears at ¾ filling. b) A near-hexagon with this orientation at ½ filling is obtained in a model of hopping to nearest- and next-nearest neighbours [$t_1$ and $t_2$ in c)] or to the next neighbour along the same bond direction [$t_1$ and $t_3$ in c), illustrated in panel b) for $t_3 = 0.15 t_1$]. Figure courtesy of V. Sunko.

The high degree of faceting of the delafossite metal Fermi surfaces leads naturally to discussion of density wave formation. However, it is important to realize that, while the ¾ filling hexagon of Fig. 5a has a nesting susceptibility to the formation of a commensurate density wave, the ½ filling hexagon of Fig. 5b is nested at an incommensurate wave-vector. This is a striking difference with the case of the square lattice, in which half filling is often the special one in terms of commensuration. This simple observation may be an important factor in the lack of observed density wave formation in the delafossite metals. Even when magnetism is observed in $PdCrO_2$, it is commensurate, and so very unlikely to be linked to Fermi surface nesting.

In the following three sections, I will expand on these introductory comments in the course of reviewing current knowledge of $PdCoO_2$, $PtCoO_2$, $PdRhO_2$ and $PdCrO_2$ in more detail.

## 5. $PdCoO_2$

*5.1 Foundational experiments*

Following the initial crystal growth and characterization of $PdCoO_2$ by the DuPont group in 1971, the next significant experimental activity was the single crystal growth and experiments by Tanaka, Takei and co-workers in Tokyo, starting in 2006 [9,10,34–44]. It is now known that these crystals were not of optimum purity, so the bulk data reported for specific heat, magnetic susceptibility and transport properties are not accurate, but the Tokyo group made a number of important observations: a) they established through angle-integrated photoemission and inverse photoemission experiments that the density of states at the Fermi level is dominated by Pd, with little or no contribution from Co [36,43,45]; b) they raised the issue that the conduction band might be formed from an *s-d* mixture of Pd states [41]; c) by studying the Knight shift they established that the intrinsic magnetic susceptibility is Pauli-like, with no evidence for intrinsic local moments [40]; d) by performing x-ray diffraction at pressures of up to 10 GPa they showed that, counter-intuitively for a layered material, $PdCoO_2$ is more compressible along its in-plane direction than along the *c*-axis, and estimated its bulk modulus to be 224 GPa [46]. They also demonstrated the feasibility of a number of substitutions and solid solutions, reporting the growth of crystals in which Mn [37] and Mg [34] were substituted for Co, and the possibility of counter-substitution of Pd and Pt [9].

The breakthrough in high purity crystal growth, arguably founding modern research on delafossite metals, was reported by Takatsu and co-workers in Kyoto in an important paper in 2007 [12]. By growing crystals with resistivity ratios of over 400, they were able to establish the bulk properties of very low-disorder $PdCoO_2$, and stimulate a larger body of work by other groups. They measured in- and out-of-plane electrical resistivity up to 500 K, as well as bulk magnetic susceptibility and specific heat. They also used Raman scattering and infrared absorption to establish the energies of prominent optical phonons, and presented an analysis of the effect of these phonons on the transport and specific heat.

When the original $PdCoO_2$ crystals were grown, the technique of Angle Resolved Photoemission Spectroscopy (ARPES) did not exist as a precision tool. Its refinement over the three decades from around 1980 has been remarkable, and, as a surface probe best suited for the study of quasi-two-dimensional materials, it is ideally suited to the study of delafossite metals. The next big step forward in the experimental understanding of $PdCoO_2$ came when Noh, Kim and collaborators grew single crystals and performed core level x-ray photoemission spectroscopy (XPS), soft x-ray absorption spectroscopy (XAS) and ARPES measurements on them [29,30,47]. The XPS and XAS results supported a picture of metallic Pd states with the Co-derived states confirmed to be $Co^{3+}$ in the low-spin state. Their ARPES data, shown in Fig. 6, were the first direct observations of the hexagonal cylinder discussed in section 4. Using a combination of slab calculations to model surface states for Pd and $CoO_2$ terminated surfaces, they identified some of the states that they observed as $CoO_2$ – derived surface states.

and then used empirical surface treatments to remove them. In this way they convincingly identified the bulk band that forms this cylinder.

Although the resolution of ARPES has been revolutionized over the past two decades, the highest resolution probe of many aspects of metallic behaviour is still the de Haas – van Alphen (dHvA) effect. It is not *k*-resolved, however, so the most powerful information comes when both ARPES and dHvA are available on the same material. Working with crystals from Kyoto, Hicks and co-workers succeeded in observing dHvA oscillations in the magnetic torque from small $PdCoO_2$ crystals mounted on piezoresistive levers [14]. They observed two closely spaced oscillation frequencies of 29.3 and 30.7 kT, with the splitting associated with the slight three-dimensional warping of the cylindrical Fermi surface. Taking the average of the two as the two-dimensional Fermi surface area gives a value of 50.1 % of the Brillouin zone, as expected for a single half-filled band, and an average effective mass of 1.5 $m_e$, allowing the precise deduction of the parameters listed in Table 2.

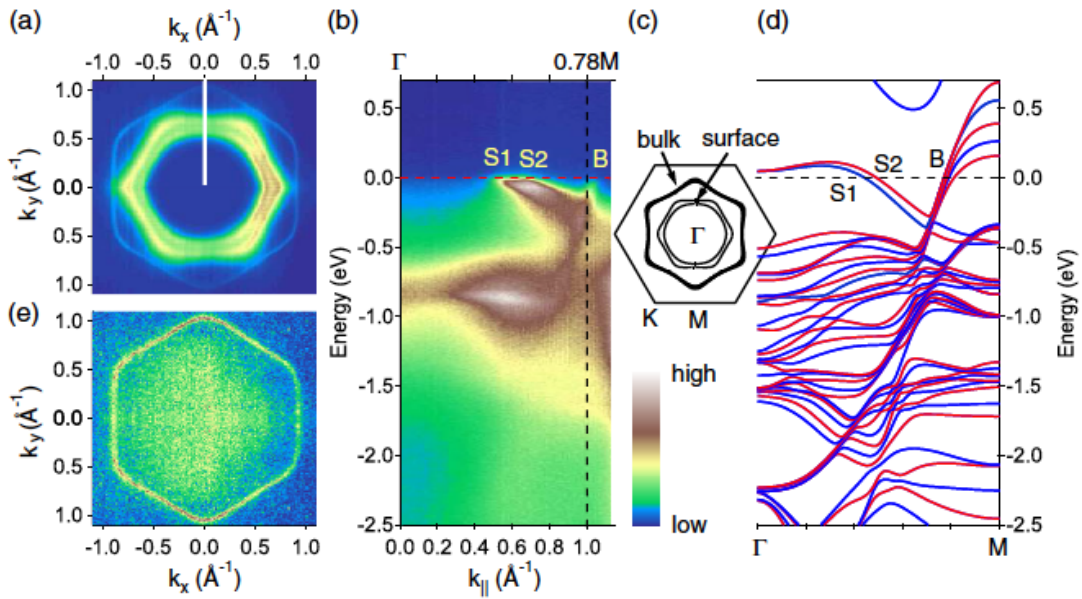

**Fig. 6** (colour online) The first published angle resolved photoemission Fermi surface of a delafossite metal, in this case $PdCoO_2$, from Noh *et al.* [29]. Data from a pristine surface show both bulk and surface states (a,b), as sketched in (c). Calculations were used to identify the surface states [labeled S1 and S2 in (d)], and the introduction of slight surface disorder was then used to destroy them, leaving the single bulk Fermi surface shown in (e).

Hicks *et al* also reported the first high-resolution studies of the low temperature resistivity of PdCoO$_2$. Their low temperature data are shown in Fig. 7. Although nominally a simple experiment, there are several pitfalls in measuring the in-plane resistivity $\rho_{ab}$ of a layered material with such a high in-plane conductivity. The first is that considerable care must be taken to ensure homogeneous current paths through the sample. If current is injected through a top surface, there is a danger that homogeneous flow will not have been obtained as it passes the points at which the voltage is measured, especially because the crystals typically have sub-mm dimensions. If the flow is not homogeneous, the current will be biased towards the upper layers of the sample. Since the temperature dependences of $\rho_{ab}$ and $\rho_c$ are quite similar [12], there will not be an obvious

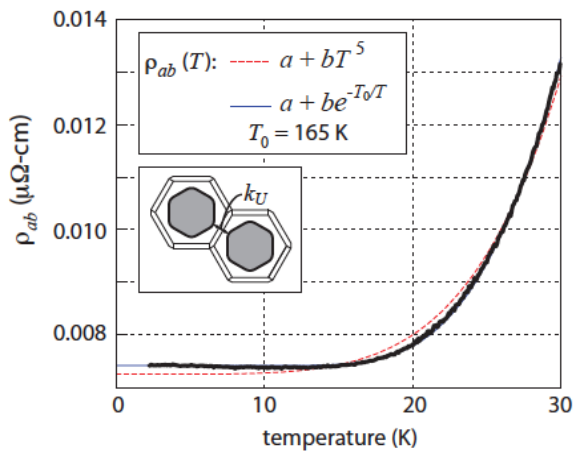

**Fig. 7** (colour online) Low temperature resistivity of PdCoO$_2$, from Hicks *et al.* [14]. The dotted red line is a fit to the functional form of the standard Bloch-Grüneisen expression, and is seen to be a significantly worse fit to the data than the exponential expression given below in the top inset. The bottom inset shows the minimum *k*-space distance between PdCoO$_2$ Fermi surfaces in adjacent Brillouin zones, which, in combination with the acoustic phonon velocity, determines the parameter $T_0$.

temperature-dependent indicator in the measured resistance that the current is not homogeneous; the main result will be a reduction in the effective electronic thickness, causing an overestimate of the value of $\rho_{ab}$. The values of the room temperature $\rho_{ab}$ reported in the modern literature show considerable variation, from 2.6 μΩcm [14] to 6.9 μΩcm [12]. Since this is measured at 300 K there is no plausible intrinsic reason for such a range, because at these temperatures the impurity contribution to the resistivity is tiny. Current path inhomogeneity is therefore the most likely cause. Special care was paid to current homogeneity and sample geometry in the work of Hicks *et al.*, so their value is quoted as intrinsic in this paper. For the low temperature properties to be studied with precision, there is a further challenge. Using typical single crystal geometries leads to low temperature resistances of 5-50 μΩ, giving a signal of only 5-50 nV at a typical measurement current of 1 mA. Noise levels of less than 0.5 nV/√Hz are therefore needed for precision measurement. These are available only in specialized cryostats; for the data shown in Fig. 7, low temperature amplification was employed to achieve a noise level of 0.15 nV/√Hz.

The data of Fig. 7 show not just a remarkably low value for $\rho_{ab}$ but also an unusual temperature dependence. The precision of the data allow a distinction to be made between the $T^5$ behavior of the standard Bloch-Grüneisen expression and an exponential increase. This probably results from an effect discussed in the 1930s by Peierls [48] and expanded on by Herring [49], Ziman [50] and others. One of the central assumptions of Bloch-Grüneisen theory is that electrons scatter from a distribution of phonons that remains in equilibrium, with zero net momentum. Normal electron-phonon scattering processes within the same Brillouin zone therefore relax electron momentum and lead to resistivity. It is difficult to see why this should be the case, particularly at low temperatures. Assuming that phonon scattering is the main source of temperature-dependent resistivity, if the phonons are being dragged out of equilibrium and gaining momentum from the electron distribution, the normal electron-phonon processes will not contribute to the electrical resistivity. The first electron-phonon processes that will relax momentum and affect the resistivity will be electron-phonon Umklapp processes in which a phonon scatters an electron to the next Brillouin zone. In a material with a single Fermi surface sheet that does not touch the Brillouin zone boundary, there is an energy gap for such a process, and an exponential resistivity is expected. Although the above arguments are persuasive, a number of complications that affect most metals led to great difficulty in observing exponential resistivity (see, for example [51]), and the arguments of Peierls and Herring have, for the most part, been quietly ignored or forgotten. It seems that the extremely low impurity scattering and the large energy gap of approximately 170 K make $PdCoO_2$ an ideal candidate for the observation of this effect.

If phonon drag is taking place, there are well-documented consequences for the behavior of thermoelectric coefficients [49,50]. One of the most prominent predictions is a peak in the Seebeck coefficient in the temperature range where the drag is most prominent. This has been observed in $PdCoO_2$ in a careful combined study of thermal and thermoelectric transport by Daou and colleagues [52], who employed finite element calculations to model electrical and heat current distributions in their samples, so a qualitatively self-consistent picture seems to have emerged.

As highlighted in the introduction, the low temperature resistivity of $PdCoO_2$ is remarkable not just for its temperature dependence but also for its absolute value of just 7.5 n$\Omega$cm (see Fig. 7). The simple single-band, closed, quasi-2D Fermi surface geometry enables the use of standard expressions to obtain an accurate estimate of the mean free path, $\ell$. As mentioned in section 2, its value of over 20 µm is extremely large, implying a huge distance between the collisions that efficiently relax momentum and lead to resistivity ([14]; see also caption to Table 2).

*5.2 Electronic structure calculations on $PdCoO_2$*

Interlinked with the above-described experimental developments were a number of electronic structure calculations [14,21,24,30,31,39,53–56]. Done with in the framework of density functional theory (DFT), these were not sensitive to the degree of correlation in the Co-derived bands, but most agreed with the basic Fermi surface topography of Fig. 6, and the dominant role of Pd-derived states in the conduction band. A topic of considerably more debate has been whether this band should be considered as derived purely from the Pd 4*d* states or is strongly influenced by the Pd 5*s* shell. At first sight, the highly faceted Fermi surface looks like it must be dominated by *d* orbitals, but the very large Fermi velocity is hard to attribute to a purely *d* band. The contribution to the DFT calculated density of states from extended orbitals is surprisingly difficult to determine, and easy to underestimate due to their interstitial weight, so this is not a question that can be answered with confidence by calculating the relative contributions to the overall density of states. In fact, Fig. 6 d) is drawn on a convenient scale to highlight the qualitative answer to this question. A very broad feature that is seen running through all the bands is the result of hybridization with the Pd 5*s* shell. Crudely speaking, the conduction band of $PdCoO_2$ might be thought to have its strong dispersion because of its 5*s* character and its pronounced faceting because of its 4*d* character. Detailed analysis of the precise interlayer dispersion that can be probed by the dHvA effect in tilted fields also led Hicks and co-workers to this conclusion, independent of electronic structure calculations [14].

Although DFT calculations are not suitable to address the degree of correlation on the Co sites directly, some indirect information could be obtained from studying the effect of including on-site repulsion *U* on the Co sites on fine details of the Fermi surface dispersion in LDA + U calculations. It was found that if the Co-based bands are treated in the complete absence of correlation, a small Co contribution to the conduction band exists in parts of the Brillouin zone. Including a *U* on the Co site reduces that coupling, and improves the detailed agreement with the dHvA data [14]. This is an interesting example of the value of these approaches. For the *U* = 5 eV adopted in [14] one would not expect the LDA + U calculations to correctly capture the physics of the Co-based bands themselves, but the secondary effect of *U* on the conduction band hybridization is more trustable.

In addition to establishing the basic band structure, the DFT-based calculations have also estimated key physical properties such as average Fermi velocities for in-and out-of-plane transport [21,31]. The numerical values obtained differ from the experimentally determined ones of Table 2, but the trends in resistivity are correct and, impressively, a successful prediction was made of the positive in-plane thermopower (counterintuitive for a single electron-like band) subsequently observed in experiments on single crystals [52]. Interesting calculations of phonon frequencies and strain-dependent physical properties have also been reported [24,54].

| | |
|---|---|
| $\rho_{ab}$ (300 K) | 2.6 µΩcm |
| $\rho_{ab}$ (4 K) | 0.0075 µΩcm |
| $\rho_{c}$ (300 K) | 1.07 mΩcm |
| $\rho_{c}$ (4 K) | 8.1 µΩcm |
| $\gamma_0$ | 1.3 mJ/molK$^2$ |
| $\chi_0$ | 2.4 x 10$^{-4}$ e.m.u./mol |
| $k_F$ | 0.96 x 10$^{10}$ m$^{-1}$ |
| $m^*$ | 1.49 $m_e$ |
| $v_F$ | 7.5 x 10$^5$ ms$^{-1}$ |
| $v_c$ | 2.3 x 10$^4$ ms$^{-1}$ |
| $n$ | 2.45 x 10$^{22}$ cm$^{-3}$ |
| $n_{2D}$ | 1.45 x 10$^{15}$ cm$^{-2}$ |
| $\rho_{2D}$(4 K) | 0.126 Ω |
| $\ell$ | 21.4 µm |

**Table 2** Summary of key metallic parameters for PdCoO$_2$. $P_{ab}$ and $\rho_c$ are the resistivity in-plane and perpendicular to the plane, from [14]. $\gamma_0$ and $\chi_0$ are the electronic specific heat coefficient and Pauli susceptibility, from [12]. Fermi vector $k_F$, effective mass $m^*$ and Fermi velocity $v_F$ are Fermi surface averages deduced from de Haas-van Alphen data [14], $v_c \equiv v_F[\rho_{ab}(4\ K)/\rho_c(4\ K)]^{1/2}$, and $n$ and $n_{2D}$ are carrier densities in three and two dimensions respectively, estimated assuming stoichiometry and monovalent Pd (which is confirmed by de Haas-van Alphen measurements of the Fermi volume). The measured Hall coefficient is negative, as expected for an electron-like Fermi surface [57]. The resistance per plane is denoted $\rho_{2D}$. Mean free path $\ell$ is calculated using the standard two dimensional expression $\rho_{ab}^{-1} = \frac{e^2}{hd} k_F \ell,$ where $h$ is Planck's constant, $e$ the electronic charge and $d$ the interlayer spacing.

*5.3 Unusual properties resulting from the extremely high conductivity of PdCoO$_2$*

The combination of unusually high conductivity and strong two-dimensionality found in PdCoO$_2$ is so unusual that it provides unique experimental opportunities to investigate new regimes of electrical transport. Since it is not yet a very well-known material, these are far from being fully exploited, but there have already been several intriguing experiments. Takatsu and colleagues [58] studied the effect of in-plane magnetic fields on the *c*-axis magnetoresistance (MR), and observed a huge 35000% MR for fields applied along the [1$\bar{1}$0] direction but a much smaller effect for fields along [110]. This is clearly related to the *c*-axis dispersion and Fermi surface topography, and the authors present a Boltzmann equation analysis based on a tight-binding parameterization of the conduction band that qualitatively reproduces the observations. The scale of the effect rests on the combination of low dimensionality, field-tuned interlayer coherence and extremely large relaxation times associated with mean free paths of the scale shown in Table 2. Importantly, the authors also established that the MR in PdCoO$_2$ obeys Kohler's rule, suggesting that all relevant scattering has its origin in a single process.

Following up on investigation of the *c*-axis MR, Kikugawa *et al*. [59] studied the effect of rotation the applied field not within the plane, but from the *c*-axis (001) to the in-plane (1$\bar{1}$0) and (110) directions. They observed the strong MR peaks expected when tuning a nearly two dimensional metal to so-called Yamaji angles at which all quasiclassical orbits have the same area, but the effect is astonishingly large – as high as 5.5 x 10$^4$ % at one Yamaji angle by their peak field of 35 T. This scale is attributable to the extremely long mean free path of PdCoO$_2$.

Motivated by recent developments in topological semi-metals that are beyond the scope of the current review, Goswami and collaborators [60] have pointed out that the axial anomaly physics that should be accessible in those materials can also in principle be studied in high purity layered metals. In a semimetal, the extreme quantum limit of all carriers in a single Landau level can be achieved in laboratory-accessible magnetic fields because the very low carrier concentration enables the confinement of all carriers in the lowest Landau level. At first sight, this is not achievable in normal metals because it would require applied fields of thousands of tesla. In a layered metal, however, if the cyclotron energy exceeds the effective interlayer hopping parameter, the Fermi level can be made to intersect only one Landau level at a time. One is not in the lowest Landau level, but one can mimic physics that requires single Landau level occupancy. Even in strongly two-dimensional materials, this normally requires fields of tens of tesla, but tilting the field to a Yamaji angle can reduce the effective hopping to zero. If this can be done in a material whose disorder scattering rate is considerably less than the cyclotron energy, new physics is in principle accessible. Kikugawa *et al.* [59] used $PdCoO_2$ to access this regime. They attributed the magnetoresistive behavior that they then observed in the vicinity of Yamaji angles to the physics of the axial anomaly, and argued that axial anomaly physics is also seen for field and current along the crystallographic *c* axis, due to the combined contributions from more than a single Landau level. As highlighted in Ref. [61], study of anistropic magnetoresistance has to be done with great care to avoid unwanted contributions from inhomogeneous current paths. The geometry used by Kikugawa *et al.* differs from that commonly used in three-dimensional Weyl and Dirac metals [61] but it would be interesting in future to see the current path homogeneity in studies of c-axis resistivity in the delafossites checked with multi-contact measurements.

Rather than concentrate on the out-of-plane properties, Moll and collaborators [62] decided to investigate the consequences of the extremely long mean-free path and phonon drag on in-plane transport. Motivated by fifty-year old theoretical work by Gurzhi [63], and experiments and theory from the 1990s on high mobility two dimensional electron gases by Molenkamp and de Jongh [64,65], they used a focused ion beam to fabricate restricted width conduction channels and study how the restriction affected the device resistance. The key idea behind the experiments is that if one can find a material in which the rate of collisions that conserve the conduction electrons' momentum is higher than the rate of collisions that relaxes the momentum, it is possible to enter a regime in which the viscosity of the electronic fluid influences the electrical transport properties. In almost all known materials this regime is unattainable for two reasons: strong disorder scattering, which transfers the fluid's momentum to the host lattice, and too few collision processes that conserve the fluid momentum. The hope was that in PdCoO$_2$, the rate of momentum-conserving processes would be increased due to phonon drag and have a chance of overwhelming the low rate of momentum-relaxing collisions indicated by the extraordinarily long mean free path. This effect was indeed observed, and the experiments enabled the first estimate of the viscosity of the electron fluid of a metal [62].

## 6. PtCoO$_2$ and PdRhO$_2$

Given the unusual properties of PdCoO$_2$, it is natural to seek related materials in which either element A or element B in the ABO$_2$ structure can be changed in such a way as to perturb those properties but not change them completely. In the delafossites, the two known opportunities to do this, i.e. to retain conductivity based on A element band(s) in non-magnetic metals, are to exchange either Pd or Co for other elements in the same column of the periodic table. The chemically known variants for which the relevant materials crystallize, both reported in the original Dupont group papers, are PtCoO$_2$ and PdRhO$_2$. Neither have been investigated so far to anywhere near the extent of PdCoO$_2$, but both offer interesting opportunities: PtCoO$_2$ to move from 4$d$ – 5$s$ conduction band physics to 5$d$ – 6$s$ conduction band physics (likely increasing the conduction bandwidth) and to step up the role played by spin-orbit coupling, and PdRhO$_2$ to increase the Pd-Pd bond lengths and thereby likely decrease the conduction bandwidth. The progress to date in investigating them will be reviewed in the following two sub-sections.

*6.1 Experiments on PtCoO$_2$*

Of all the materials synthesized by Shannon and collaborators in their foundational work of the 1970s, PtCoO$_2$ arguably interested them the most, because of the rarity of observing Pt in an apparent 1+ oxidation state. They successfully grew single crystals, and established that they had a room temperature resistivity of approximately 3 μΩcm and anisotropy of approximately 300, as mentioned in Section 1.2 [1–3]. They later used reactive sputtering to produce thin films [66] that were used by another group for Raman experiments [67] but there seem to be no reports of follow-up measurements on the crystals.

As with $PdCoO_2$, the next generation of crystal growth and experiments came from Tanaka and colleagues at the University of Tokyo, but in the case of $PtCoO_2$, the as-grown crystals were very small (only 30 μm x 30 μm) so few new single crystal experiments were performed. Powdered crystals were reported to have a strong Curie-Weiss contribution to the magnetic susceptibility [9], but follow-up NMR experiments of $^{195}$Pt and $^{59}$Co correctly established that the previous observation was due to impurity phases, and that the intrinsic local internal magnetic susceptibility was temperature independent and Pauli-like [40]. Ref. [68] reported in-plane resistivity of 0.55 μΩcm at 16 K and 4.69 μΩcm at 260 K, but neither further data nor the geometry of the crystal from which these values were obtained seem to have been reported. In the same paper, Higuchi *et al.* used resonant photoemission spectroscopy (angle-integrated) and x-ray absorption spectroscopy to establish that the states at the Fermi level were dominantly Pt 5*d* character, and stated that the conduction band was probably the result of 5*d*-6*s* hybridization [68].

The first growth of $PtCoO_2$ single crystals with dimensions of several hundred μm since the work of the Dupont group was reported by Kushwaha and colleagues in a preliminary manuscript in 2014 [69] and a full paper including angle resolved photoemission spectroscopy in 2015 [15]. The Fermi surface and quasiparticle dispersion of $PtCoO_2$ from those photoemission experiments are shown in Figs. 8 and 9. In combination with de Haas – van Alphen measurements (cited in Ref. [15] but not yet published in detail) they confirm the Fermi surface averaged mass to be 1.18 $m_e$, and the average Fermi velocity to be 8.9 x $10^5$ ms$^{-1}$, approximately 20% lower (higher) than those shown in Table 2 for PdCoO$_2$. The conduction band is electron-like, and, as for PdCoO$_2$ [57] the Hall resistivity is negative [15].

Focused ion beam sculpting was used to define precise crystal geometry and establish that the room temperature resistivity is 2.1 μΩcm, falling to ~ 0.04 μΩcm at a resistive minimum seen at 16 K. Further work will be required to establish whether that minimum is intrinsic or dependent, for example, on the width of the conducting channel, but the room temperature value should be accurate to within 5%. Although not as long as that of PdCoO$_2$, the mean free path at the resistivity minimum is still extremely long by the standards of metallic oxides, at approximately 4 μm. A further report of the study of the *c*-axis transport of PtCoO$_2$ single crystals appears in [59] but the only report is of c-axis magnetoresistance normalized to resistance in zero applied field; no details are given of crystal growth or of transport data in absolute units. The quantities for which reliable data exist in absolute units are summarized in Table 3.

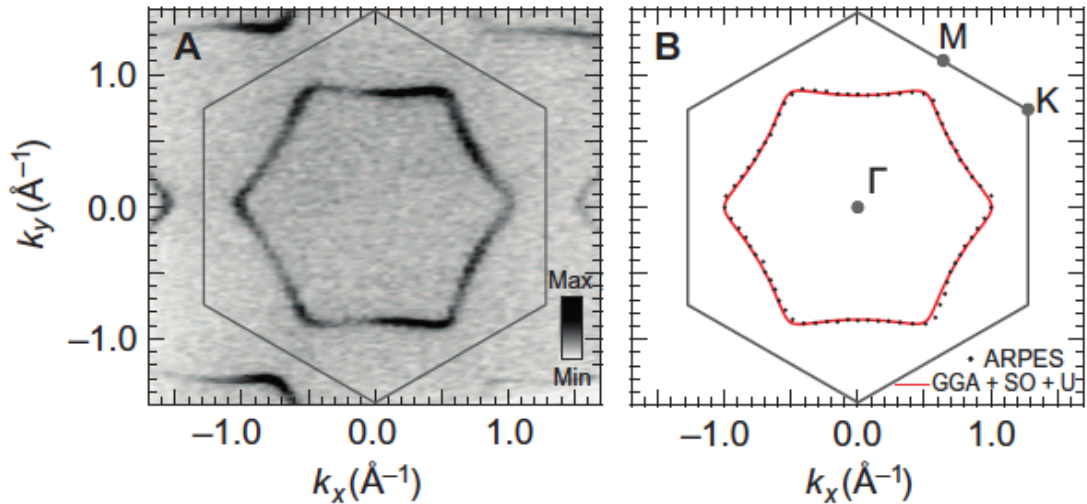

**Fig. 8** (colour online) The Fermi surface of the bulk states in PtCoO$_2$ from raw photoemission data (A) and fitted spectra (B). The red line in B is the prediction of an electronic structure calculation including spin-orbit coupling and a realistic value of *U* on the Co sites, with its area scaled by 8% to match the experimental observation (see main text). From Kushwaha *et al.* [15].

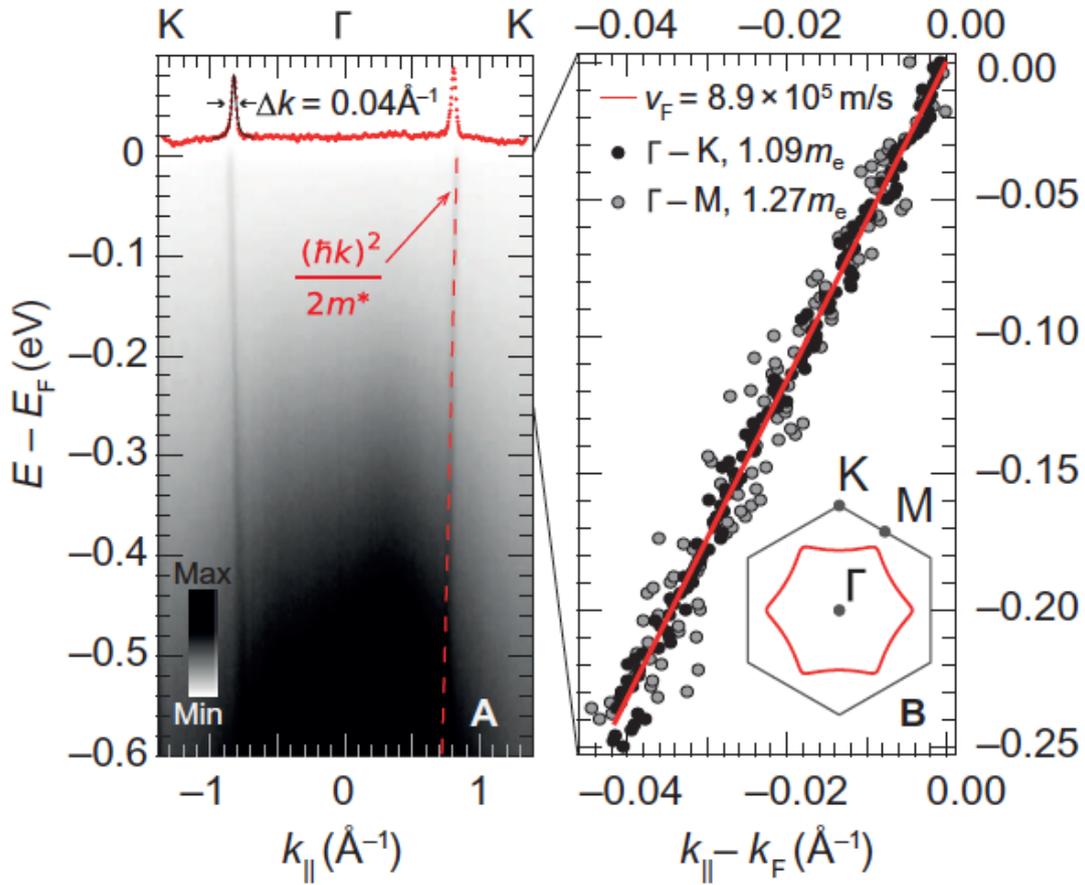

**Fig. 9** (colour online) Photoemission data for the energy-wavevector dispersion of $PtCoO_2$, as raw data (A) and as obtained from fitting of spectra (B). A linear fit in the vicinity of the Fermi level established the extremely high Fermi velocity. From Kushwaha *et al.* [15].

*6.2 Electronic structure calculations on $PtCoO_2$*

Early electronic structure calculations were reported in refs. [39,55,56], but these concentrated mostly on the density of states and its orbital decomposition, with the calculations on $PtCoO_2$ not being the main focus of the papers. More comprehensive results were reported by Ong and collaborators in [21] and [31], who showed that, if spin-orbit coupling is taken into account, the calculated Fermi surface is the single cylinder of hexagonal cross-section characteristic of the delafossite metals. The importance of spin-orbit coupling in creating this agreement was stressed in Ref. [15]. Calculations that do not incorporate spin-orbit coupling predict that a second band crosses the Fermi level, so spin-orbit coupling plays an important role in maintaining the appealingly simple single band Fermi surface of PdCoO$_2$ in PtCoO$_2$. In line with the experimental results, a larger overall conduction bandwidth is predicted for PtCoO$_2$. As shown in Fig. 8, excellent agreement is obtained between theory and experiment for the detailed shape of the Fermi surface, in calculations incorporating spin-orbit coupling and including a realistic $U$ on the Co sites. PtCoO$_2$ was included in the survey calculation of structural properties by Kumar and collaborators [24], and electronic structure calculations have been used to predict a number of properties that have not yet been studied experimentally. Ong and colleagues emphasized the possibility of a very large *c*-axis thermopower [31], a theme expanded on in calculations by Gruner and colleagues, who argued that in-plain biaxial strain could further enhance the thermoelectric figure of merit in $PtCoO_2$ for *c*-axis transport [54]. In this paper, $PdCoO_2$ and $PtCoO_2$ were compared, and $PtCoO_2$ predicted to be the more promising thermoelectric material. The strains required were predicted to be large, however (several per cent) so it is unclear how practical it would be to achieve them in reality.

| | |
|---|---|
| $\rho_{ab}$ (300 K) | 2.1 μΩcm |
| $\rho_{ab}$ (4 K) | 0.04 μΩcm |
| $k_F$ | 0.95 x 10$^{10}$ m$^{-1}$ |
| $m^*$ | 1.14 $m_e$ |
| $v_F$ | 8.9 x 10$^5$ ms$^{-1}$ |
| $n$ | 2.42 x 10$^{22}$ cm$^{-3}$ |
| $n_{2D}$ | 1.44 x 10$^{15}$ cm$^{-2}$ |
| $\rho_{2D}$(4 K) | 0.67 Ω |
| $\ell$ | 4 μm |

**Table 3** Summary of known metallic parameters for PtCoO$_2$. $\rho_{ab}$ is the resistivity in-plane, from [15]. Fermi vector $k_F$, effective mass $m^*$ and Fermi velocity $v_F$ are Fermi surface averages deduced from de Haas-van Alphen data [15] and $n$ and $n_{2D}$ are carrier densities in three and two dimensions respectively. The 2% deviation from monovalent filling reported from de Haas-van Alphen measurements has been included in the estimates of $k_F$, $n$ and $n_{2D}$, but further work will be required to determine whether this deviation is real or due to experimental error. The measured Hall coefficient is negative, as expected for an electron-like Fermi surface [15,69]. The resistance per plane is denoted $\rho_{2D}$. Mean free path $\ell$ is calculated using the standard two dimensional expression $\rho_{ab}^{-1} = \frac{e^2}{hd} k_F \ell$, where $h$ is Planck's constant, $e$ the electronic charge and $d$ the interlayer spacing.

*6.3 Stoichiometry of $PtCoO_2$*

Starting from the first synthesis of $PtCoO_2$, questions have been raised about intrinsic non-stoichiometry [2], and literature reports exist of strongly non-stoichiometric crystals. Chemical analysis can often contain larger systematic errors than expected, however, so it is useful to have an independent check. In a single-band material like $PtCoO_2$, measurement of the Fermi volume is, in principle, just such a cross-check. The formal valences $Pt^{1+}Co^{3+}O_2^{2-}$ mean that unless there is a compensation between cation and oxygen non-stoichiometry, vacancies of Pt, Co or both, or any cross-substitution between Pt and Co (the latter chemically unlikely) would have a direct effect on the Fermi level, and thus affect the Fermi volume. One can therefore use advanced experimental physics techniques to gain essentially chemical information.

At first sight, photoemission is an excellent way of measuring the Fermi area (which is easily converted to a Fermi volume in a nearly two-dimensional metal), but the cleaved surfaces of the delafossites are polar. The fact that they support strong surface states confirms that this polarity is important, and that significant surface and near-surface electric fields are present. That in turn means that deviations from half filling of a few per cent observed in photoemission should not immediately be taken as proof of non-stoichiometry, because there is the possibility of surface-dependent changes in the Fermi level. Indeed, in $PtCoO_2$, the photoemission Fermi surface shown in Fig. 9 is approximately 8% smaller than the value expected at half-filling. A more reliable probe is therefore the de Haas-van Alphen effect, which is sensitive to the true bulk. In the crystals studied in ref. [15], the Fermi surface de Haas-van Alphen derived Fermi surface area is reported to be within 2% of the value expected for a stoichiometric material. This is beyond the expected uncertainties of the technique, so may indicate a slight non-stoichiometry, especially since equivalent checks in $PdCoO_2$ imply stoichiometry to within experimental error. However, even a 2% non-stoichiometry would be incompatible with a 4 μm mean free path in any standard metal, so more work will be required in order to understand this important point. It would be particularly interesting to see if de Haas-van Alphen frequencies vary from crystal to crystal, and if that variation could be correlated with measurements of the mean free path. Already, however, enough is known about the properties of $PtCoO_2$ single crystals to effectively rule out the large deviations from stoichiometry suggested by chemical analyses presented in some of the early papers.

*6.4 Information to date on $PdRhO_2$*

$PdRhO_2$ is by some distance the least studied of the delafossite metals. Although synthesized by the Dupont group, it was reported only as a black powder, and although they later grew sputtered thin films [66], there have been no subsequent reports in the literature of single crystal growth. Its Raman-active phonons were subsequently observed in thin films produced by reactive sputtering and the oxidation of alloy foils [70] but no other experimental work has been reported so far. Calculations do exist, and give a strong motivation for future experiments. As well as having similar lattice excitations [24], $PdRhO_2$ is predicted to have the same basic single-band nature as $PdCoO_2$ and $PtCoO_2$, and a similar nearly hexagonal Fermi surface cross-section [22]. However, its substantially larger lattice parameters lead to a reduction of bandwidth, offering a valuable opportunity to test understanding gained from $PdCoO_2$ and $PtCoO_2$ in a slightly different environment.

## 7. A magnetic delafossite metal: PdCrO$_2$

*7.1 Spin structure, metallic conduction and coupling between magnetism and conduction electrons*

In the original Dupont group paper on delafossites, relatively little was said about PdCrO$_2$, which was not grown in single crystal form and described only as a 'black powder'. In 1986, comparison of its magnetic susceptibility with those of a series of insulating delafossites led Doumerc *et al.* [71] to the conclusion that it was an antiferromagnet with a Weiss temperature of approximately -500 K and an effective moment of approximately 4 Bohr magnetons per Cr. The hypothesis of antiferromagnetism was confirmed by a neutron scattering experiment a decade later by Mekata *et al.* [72], who showed that it is based on a commensurate 120° spin structure. Takatsu and collaborators performed a thorough investigation of polycrystals of PdCrO$_2$ in 2008-9 [73,74]. Building on the earlier findings, they confirmed a 120° spin structure and established further key facts: a) PdCrO$_2$ is metallic, and has a Néel temperature of 37.5 K at which a feature is seen in the resistivity, i.e. there is coupling between the conduction electrons and the magnetism; b) in line with its large frustration parameter of 13, analysis of the specific heat shows that less than 1/3 of the expected magnetic entropy expected for Cr$^{3+}$ is released at $T_N$, interpreted by the authors as evidence for short-range spin-spin correlations persisting to much higher temperatures. This conclusion was also supported by a scaling analysis of the specific heat, which showed nearly temperature-independent critical exponents over an unusually wide temperature range. A brief subsequent pressure study has indicated that the basic features established about the magnetism of PdCrO$_2$ in Refs. [73,74] are pressure-independent in modest applied pressures of up to 10 kbar [75].

The next breakthrough came from the same group. Motivated by the possibility of high-resolution study of metallic behavior in the presence of frustrated magnetism, Takatsu and Maeno [13,76] grew the first reported single crystals of PdCrO$_2$. Like the non-magnetic metals, they were very pure, with residual resistivities of 0.045 μΩcm, but the room temperature resistivity of approximately 8 μΩcm is larger than that of $PdCoO_2$. Takatsu and colleagues attributed this to extra magnetic scattering due to the fluctuating $Cr^{3+}$ moments, and indeed used direct subtraction of the $PdCoO_2$ data to isolate the magnetic part of the resistivity of $PdCrO_2$.

The effects of coupling of $CrO_2$ layer magnetism to the conduction electrons has subsequently been studied by a number of groups. Several specific issues surrounding this coupling are discussed in separate sections below, but it is clear from measurements of Raman scattering and electron spin resonance [25] and magnetothermopower [78] that it influences a wide range of physical properties over a wide range of temperatures, extending to well above $T_N$.

*7.2 Possibility of chiral magnetism and unconventional Hall effect*

In growing the single crystals, one of the goals of Takatsu and collaborators was to study the Hall effect in the presence of the $CrO_2$ layer magnetism, motivated by recent theoretical advances linking chiral magnetism with a novel contribution to the Hall voltage in what is often referred to as the unconventional anomalous Hall effect, or UAHE. A full description of this subtle effect, which is related to the behavior of the Berry phase in certain magnetic structures, is beyond the scope of this article; for a recent review the reader is referred to [79] and references therein. Repeating the procedure of subtraction of $PdCoO_2$ data to isolate the magnetic component of the Hall response, they reported the observation shown in Fig. 10, in which they plot $R_S$, the magnetic component of the Hall response normalized to the measured magnetization. In the conventional anomalous Hall effect, the Hall response scales with the magnetization, so $R_S$ is expected to be field-independent. The clear deviation from this field independence at low temperatures is associated by the authors with the unconventional contribution [57].

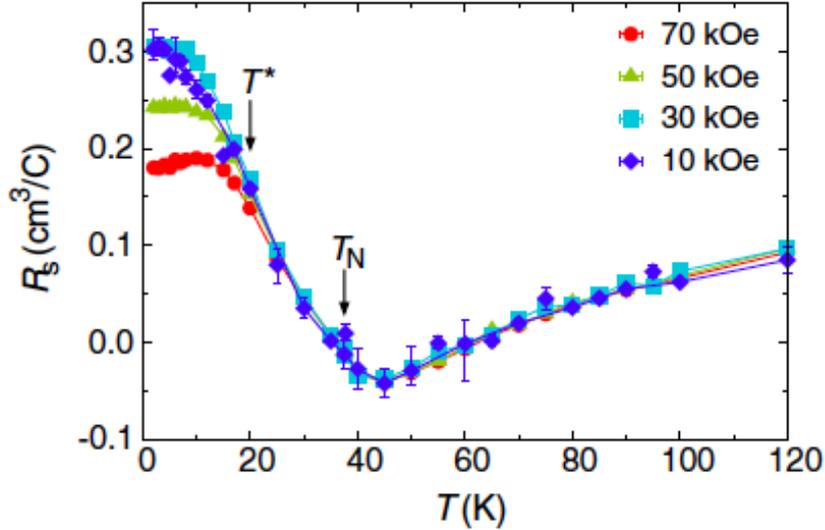

**Fig. 10** (colour online) The magnetic component of the Hall voltage of $PdCrO_2$, normalized to its measured magnetization. The noteworthy feature is the pronounced field dependence at low temperatures; see main text for discussion. Figure from Takatsu *et al.* [57].

The findings of Ref. [57] raised a number of important questions, and stimulated considerable further work. Although the technique of subtracting data from non-magnetic $PdCoO_2$ is in principle a neat way of isolating the magnetic contribution in $PdCrO_2$, it relies on the assumptions the Fermi surfaces of the two materials are essentially identical above $T_N$, and very similar below the magnetic transition. Also, for the low temperature behavior to be ascribable to the unconventional anomalous Hall effect, it is not enough for the magnetic structure to be simple 120° antiferromagnetism because that does not automatically incorporate the necessary chiral spin structure. Indeed, alternative interpretations of the findings of Ref. [57] were subsequently proposed in Refs. [80] and [32], as will be discussed further in section 7.3. Finally, for the non-magnetic subtraction to be valid at high temperatures, one needs to be sure that there are still fluctuating moments at these temperatures, i.e. that Cr stays in its magnetic 3+ configuration well above $T_N$. All of these questions were difficult to answer from first principles by calculation (though a number of trial magnetic structures were compared and contrasted in Ref. [81]), and instead required further experiments.

By far the most thorough magnetic structure measurements yet performed on PdCrO$_2$ were reported in Ref. [82]. The authors conclude that the spin structure is either that of a commensurate 120° antiferromagnet in which the spin plane includes the *c* axis and the orientation of the spins changes periodically from layer to layer, or a slightly more complicated structure in which the spin plane itself also shows layer to layer modulation. The distinction is important in terms of the discussion of the UAHE, because only the latter structure includes the chirality required to validate the interpretation of the low temperature data shown in Fig. 10 in terms of the UAHE. On the basis of the currently available experimental evidence it is not possible to distinguish the two possible spin structures at low temperatures, so the question remains open.

*7.3 Fermi surface of PdCrO$_2$ above and below T$_N$*

As with non-magnetic PdCoO$_2$ and PtCoO$_2$, the strong two-dimensionality of the electronic structure and high purity of PdCrO$_2$ make it ideal for study by both photoemission and the de Haas – van Alphen effect, whose combination has proved to be particularly powerful in understanding its electronic structure. The first reported experiment was an ARPES study by Sobota and colleagues [33]. They successfully distinguished surface and bulk states using both calculations and thermal cycling to destroy the surface states, and showed that the bulk Fermi surface of PdCrO$_2$ at 50 K is almost identical to that of non-magnetic PdCoO$_2$. The surface states observed were from Pd-terminated surfaces, also very similar to those predicted for PdCoO$_2$ [30]. However, as discussed extensively in their paper, the surprise was that the observed Fermi surface at 10K was essentially identical to that at 50K, implying that the antiferromagnetic transition at 37.5 K had no observable effect on the conduction electrons. As discussed above, the resistivity clearly indicates coupling between the magnetic and conduction sub-systems, so Fermi surface gapping at the antiferromagnetic wavevectors was expected.

The expected Fermi surface gapping was indeed observed in a de Haas-van Alphen experiment by Ok and collaborators [80], published almost simultaneously with the ARPES results of Sobota *et al*. Combining analysis of the dHvA frequencies with a band structure calculation in the presence of 120° magnetic order, Ok *et al.* showed that their results were qualitatively consistent with the opening of very small gaps at the boundaries of the new magnetic Brillouin zone. By performing measurements in high magnetic fields of up to 33 tesla, they showed that these gaps were present not only at the very low temperatures to which dHvA is usually restricted, but that they persisted to approximately 25 K [80]. At low temperatures, they saw both low frequency oscillations from the small Fermi surface sheets produced by the gapping and high frequency oscillations from orbits resulting from field-induced tunneling across the gaps (see Fig. 11), and estimated a very low magnetic breakdown field of approximately 7 T.

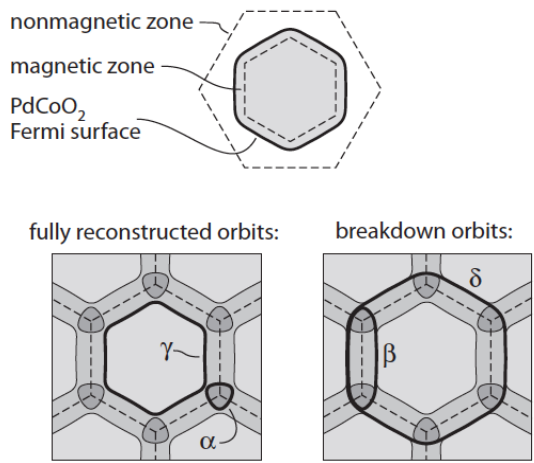

**Fig 11** Illustration of the Fermi surface and expected de Haas-van Alphen orbits after a PdCoO$_2$-like high temperature Fermi surface is folded into the magnetic Brillouin zone created by the antiferromagnetic order in PdCrO$_2$. The α and γ orbits are seen in the presence of small gapping due to the interaction of the conduction electrons with the localised magnetism, while breakdown orbits β and δ are observed when the field is high enough to induce tunneling across those gaps. Figure from Hicks *et al.* [77].

Using the same in-house grown single crystals as those studied in Ref. [80], Noh and collaborators [32] performed new photoemission experiments in search of the band folding that would be expected in the presence of the gapping seen in the dHvA experiments. Working at a photon energy of 120 eV (larger than that employed in Ref. [33]), they reported the appearance of this folding for temperatures below $T_N$, as shown in Fig. 12. They also provided evidence that the magnetism was due to ordering of local moments, and stressed that this was therefore the first photoemission observation of the effect on conduction electrons of local moment magnetism.

The fourth paper in the series establishing the Fermi surface of $PdCrO_2$ below $T_N$ was that of Hicks and collaborators [77]. They reported qualitatively identical dHvA data to those of Ok *et al*, performed an extended analysis to consider Fermi surface dispersion along $k_z$, and discussed both a spin flop transition and high temperature transport properties. For the purposes of the current discussion, however, the most important aspect of their paper was their analysis of the Fermi surface folding in a purely two-dimensional approximation. Rather than comparing with electronic structure calculations, they showed that folding the $PdCoO_2$ Fermi surface into the magnetic zone (including only a very small correction to account for the different in-plane lattice parameters of $PdCoO_2$ and $PdCrO_2$) quantitatively reproduced the observed dHvA frequencies. The significance of this simple analysis is that it provides strong evidence that, above $T_N$, the non-magnetic Fermi surface is precisely that of a non-magnetic monovalent metal, i.e. one in which the $CrO_2$ layer is insulating. This is entirely consistent with the photoemission observations of Refs. [32] and [33] of the Fermi surface for temperatures higher than $T_N$. Further discussion of this point forms the basis of the following section. The fact that the low temperature Fermi surface of $PdCrO_2$ is produced by folding a hexagonal paramagnetic Fermi surface into a magnetic Brillouin zone with very small gapping also means that the ungapped Fermi surface can be used to estimate the mean free path without introducing large errors. For the observed low temperature resistivity of 0.045 µΩcm [76] this implies a mean free path of approximately 4 µm, using the known

Fermi wavector, interplane separation and the expression defined in the caption to Table 4 [3]

In closing the current section, I should return to the comments made in Refs. [32] and [80] about the interpretation of Ref. [57] that the data below 20 K shown in Fig. 11 are due to the unconventional anomalous Hall effect. Their papers appeared before the most recent magnetic structural work of Ref. [82] and their discussion is partly based on the assertion that the magnetic structure is one of the variants of simple 120° antiferromagnetism that is not intrinsically chiral, meaning that the effect observed by Takatsu and colleagues [57] must have another origin. As an alternative, they propose that breakdown across the magnetic gaps in the low temperature Fermi surface could lead to unusual field dependences of Hall effect. Ok and collaborators [80] performed further Hall effect studies near $T_N$ and also observe deviations from the conventional anomalous Hall effect. They assert that at these temperatures, the field and temperature-induced excitations from a magnetically ordered structure that is non-chiral in the ground state can have large chirality (see also Hemmida *et al.*, Ref. [83]), and argue that the UAHE is in fact observed in $PdCrO_2$ only near $T_N$. In their later paper examining candidate spin structures, however, Takatsu *et al.* state that the non-chiral variant of the two most favoured candidate structures would not have field-induced chirality at any temperature for fields parallel to the *c*-axis [82]. They argue that the intrinsically chiral alternative structure is most likely; presumably this means that the UAHE is a candidate mechanism for the observations in both temperature regimes [57]. In common with many other studies of unusual features in the Hall effect, it seems that in $PdCrO_2$ no consensus has yet been reached on the correct interpretation of the observations.

---

[3] A much larger value of 30 μm is quoted in Refs. [57,76]. The discrepancy is mainly due to a numerical/typographical error in that paper.

| | |
|---|---|
| $\rho_{ab}$ (300 K) | 8.2 µΩcm |
| $\rho_{ab}$ (4 K) | 0.045 µΩcm |
| $\gamma_0$ | 1.4 mJ/molK$^2$ |
| $\chi_0$ | 2.75 x 10$^{-3}$ e.m.u./mol |
| $k_F$ | 0.93 x 10$^{10}$ m$^{-1}$ |
| $m^*$ | 1.55 $m_e$ |
| $v_F$ | 7.5 x 10$^5$ ms$^{-1}$ |
| $n$ | 2.37 x 10$^{22}$ cm$^{-3}$ |
| $n_{2D}$ | 1.43 x 10$^{15}$ cm$^{-2}$ |
| $\rho_{2D}$(4 K) | 0.75 Ω |
| $\ell$ | 4 µm |

**Table 4** Summary of key metallic parameters for PdCrO$_2$. $P_{ab}$ is the resistivity in-plane from [77]. $\gamma_0$ and $\chi_0$ are the electronic specific heat coefficient and magnetic susceptibility measured at 4K from [74]. In this ordered magnet both contain contributions from the magnetic Cr$^{3+}$ sub-system; these have been subtracted for $\gamma_0$, but the value given for $\chi_0$ is the total susceptibility. Fermi vector $k_F$, effective mass $m^*$ and Fermi velocity $v_F$ are Fermi surface averages deduced from de Haas-van Alphen data assuming monovalent Pd [77], and $n$ and $n_{2D}$ are carrier densities in three and two dimensions respectively. All Fermi surface-related data are estimated for the electron-like Fermi surface that is produced by unfolding into the non-magnetic Brillouin zone, assuming, stoichiometry and monovalent Pd. The resistance per plane is denoted $\rho_{2D}$. Mean free path $\ell$ is calculated using the standard two dimensional expression $\rho_{ab}^{-1} = \frac{e^2}{hd}k_F\ell$, where $h$ is Planck's constant, $e$ the electronic charge and $d$ the interlayer spacing.

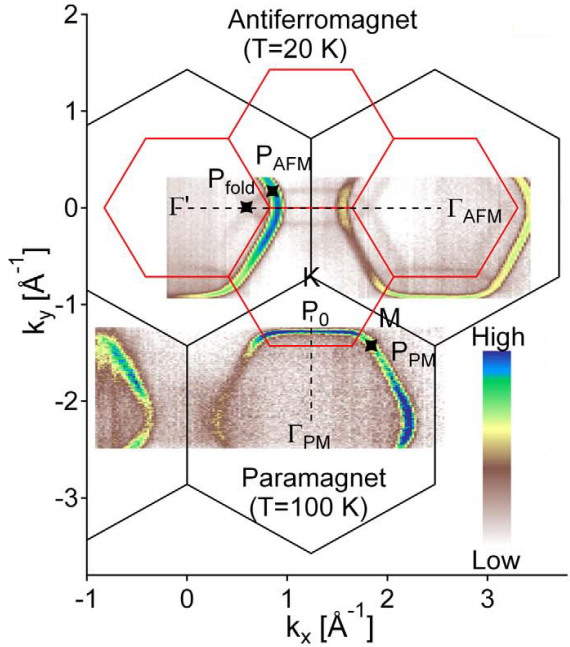

**Fig. 12** (colour online) Photoemission spectra at 100 K in the paramagnetic state of PdCrO$_2$ show a single hexagonal Fermi surface (bottom), while those at 20K show extra band folding (top). The folded bands are consistent with the opening of gaps at the antiferromagnetic zone boundaries. These gaps are too small to be resolved by the photoemission, but the new Fermi surface represented by the folded bands predicts dHvA frequencies in agreement with those observed in Ref. [80], providing good evidence that observed folding is due to the magnetic gapping. Figure from Noh *et al.* [32].

*7.4 Mott vs Slater revisited: are the Cr-derived states in PdCrO$_2$ localised or itinerant above $T_N$?*

The general issue of whether antiferromagnetism is primarily due to electron correlation or Fermi surface nesting has been discussed for decades. because Commensurate spin density modulation on a square lattice can have two origins. It is a natural consequence of electron correlation in a Mott insulator, but can also arise in the presence of weak correlations if the Fermi surface is close to that predicted in a nearest-neighbour tight binding band at half filling. One of the fathers of band theory, Slater, was a strong advocate that nesting was the dominant effect, and he and Mott had a long-running and intense debate on the issue. The discovery of a large number of Mott insulators has led to widespread acceptance that correlations are the primary driver of the phenomenon, to the extent that most papers on PdCrO$_2$ have implicitly assumed this to be the case, and used the language of local moments to describe the behavior of the CrO$_2$ layer. In a recent paper, Billington *et al.* took a different approach, and argued that above $T_N$, the Cr states are metallic (as would be predicted by simple electron counting in the absence of strong correlations), with the Fermi level such that the Fermi surface is formed from three mainly Cr-derived bands rather than a single Pd-derived band [26]. The electronic structure calculations used to reach this conclusion implicitly have to assume no large local moments, since density functional theory is not applicable to Curie-like paramagnets, and so must be of questionable validity. However, the authors performed Compton scattering experiments and used the results to construct an electron density in agreement with that predicted by the calculations.

Although the results of Billington *et al.* are intriguing, and merit further investigation, the balance of the available evidence favours the interpretation in which the CrO$_2$ layers are Mott insulators. There is persuasive evidence from magnetization and specific heat that local moments with strong local spin-spin correlations exist far above $T_N$ [74], and from photoemission that a single Fermi surface becomes gapped at the antiferromagnetic transition [32]. Billington *et al.* argue that some of the Fermi surface is not seen by photoemission for some reason, but even if one accepted that argument, the surface that *is* observed would be expected to change at $T_N$ in a way that is not observed experimentally. The de Haas-van Alphen results are all obtained below $T_N$, but can be reproduced so well by the hypothesis that a small gap opens on a single pre-existing Fermi surface that they are fully consistent with the conclusions from the photoemission. It is also important to note that the resistivity of PdCrO$_2$ falls at $T_N$, consistent with a freezing out of spin scattering but not, in a naïve picture, with a transition that gaps away half the carriers that exist above $T_c$. The final and arguably most powerful piece of evidence in favour of a localised Mott state are the x-ray absorption results of Ref. [32], which are shown in Fig. 13. The spectra both above and below $T_N$ are identical, and contain all the features of a localised $Cr^{3+}$ ion. It would be interesting to see how these data can be reconciled with the Compton scattering data of Ref. [26].

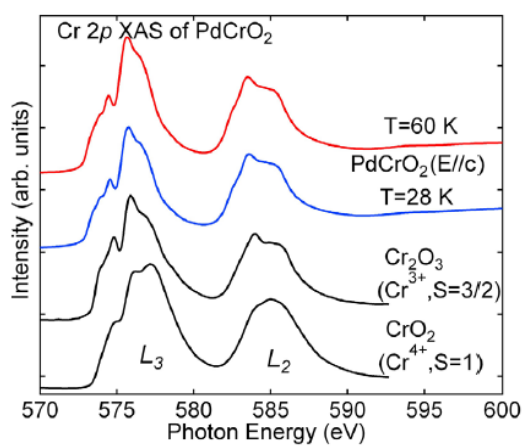

**Fig. 13** (colour online) X-ray absorption spectra for $PdCrO_2$ taken both above and below the Néel temperature, compared to those from $Cr_2O_3$, an antiferromagnetic insulator in which Co is in the 3+ state, and $CrO_2$, a metal in which Cr is in its nominal 4+ state. Figure from Noh *et al.* [32].

## 8. Summary and future work

*8.1 Brief summary*

The four points listed below give a summary of what I believe to be the most important basic properties determined so far in studies of the Pd- and Pt- based delafossite metals that I have reviewed in the main body of this article. With one exception I do not include citations, since these have been given in the main body of the review and it seems redundant to repeat them in this closing section.

1. $PdCoO_2$ and $PtCoO_2$ are established both experimentally and through electronic structure calculations to be layered metals in which a single band with a remarkably high Fermi velocity crosses the Fermi level. The dominant contribution to this band comes from Pd or Pt states, in some admixture of *d* and *s* levels. $PdRhO_2$ is predicted to have the same qualitative conduction band with a slightly narrower width, though there is as yet no experimental data to test this.

2. Although electron counting or standard band theory would not predict PdCrO$_2$ in its paramagnetic state to have a conduction band of the same character as that in PdCoO$_2$, most experimental evidence says that it does (I note that a different interpretation is proposed in Ref. [26]). These observations imply that the Cr-O layer is a Mott insulator. Local moments are present to high temperatures, with substantial short-range spin-spin correlations, which become long-range at the measured Neel temperature of 37.5 K.

3. The antiferromagnetic transition temperature of PdCrO$_2$ is strongly suppressed due to frustration since the Cr atoms in any layer sit on a triangular lattice. The full magnetic structure in the ordered state is complicated, but for a single plane is the expected 120° antiferromagnetism. Knowing the overall magnetic structure is of relevance to analysis of the Hall effect in the magnetically ordered state.

4. The Pd- and Pt- based delafossite metals have remarkably long mean free paths at room temperature (over 700 Å in PtCoO$_2$, longer than that of Cu, Ag, Pd, Pt or any alkali metal) and at low temperatures. In the best crystals, values of > 20 μm have been observed in PdCoO$_2$, 4 μm in PtCoO$_2$, and 4 μm in PdCrO$_2$. These values are very surprising in as-grown crystals.

*8.2 Future directions*

As stated in the introduction to this article, my aim in writing it was not to collate the cumulative achievements of a mature field, but to demonstrate that the foundations for a new one have been laid. I believe that the properties outlined above merit considerable future attention. Simple single-band layered metals are surprisingly rare, and ones with mean free paths as large as those of the delafossites are rarer still.

Broadly speaking I would expect future research to fall into three classes – work profiting from the properties described in this review, work aimed at understanding why the delafossite metals have these properties in the first place, i.e. what is so special about them, and work to discover and refine materials.

If one takes the approach of simply accepting the huge mean free paths and strong two-dimensionality of the delafossite metals as a welcome gift, there are a number of avenues for future research. Extension of the study of hydrodynamic effects in in-plane transport is an obvious one, as is the construction of devices on shorter length scales to test ballistic transport effects in a high carrier concentration material. There is also scope to look for new physics in which both hydrodynamic and ballistic transport regimes are present. Interlayer transport is also likely to attract considerable interest, because it might prove to be a sensitive probe of the interplay between free-electron-like and correlated electron properties that is a feature of the materials. Another intriguing limit is that in which all carriers are placed in the same Landau level. That requires high magnetic fields, strong two-dimensionality and an extremely low scattering rate. It is inaccessible in almost all known metals but might be attainable in the delafossites even without the clever use of angular tuning employed in Ref. [59]. Although that single filled Landau level would have the high level index expected of a metal, the situation would have some features in common with lowest Landau level physics in low carrier density systems, with the hope that novel collective states might be observable. Further work on the magnetism of $PdCrO_2$ and its effect on transport is also likely, as is more research on magnetothermal properties.

At least as interesting as the avenues outlined above is the quest to understand key issues such as why the delafossites have such high Fermi velocities, why the room temperature resistivity is so low, why the low temperature mean free path is so long and whether it can be made even longer. A first step to understanding the room-temperature resistivity will be to understand electron velocities, electron-phonon coupling and scattering. In this regard, several factors may come into play, alone or more likely in combination. The first is the extremely high Fermi velocity. It is not obvious that it should be so high in bands whose character contains at the least a strong *d*-electron contribution. It seems important to understand in depth the extent to which correlations on the B site layers influence the Fermi velocity of the A-site based conduction band, and more insight is also desirable into the role played by the oxygen atoms that are linearly co-ordinated with the A site atoms. In addition to the velocities being unusually high, the room temperature resistivity might be low because of unusually weak electron-phonon scattering. Detailed measurements and calculation of phonon spectra and electron-phonon coupling are highly desirable. The quasi-two-dimensionality may be playing a role, and it is also possible that if the orbital content of the states making up the Fermi surface is strongly **k**-dependent, there may be a suppression of the final states available for scattering. The interplay between surface and bulk states, and the difference between the states on A- and B-site terminated surfaces is also of considerable interest. I list all of the above in the spirit of speculation; hopefully future work allows us to identify which (if any!) of these ideas are productive.

The third likely avenue of research will be in chemistry and materials. The first hope is that existing crystals can be made even better, and that crystals are grown of materials for which none have been reported. The broader hope is that if the properties of the delafossites can be better understood, new materials can then be designed. Part of this work might be conventional chemistry, but there seems to be plenty of uncharted territory in thin film growth, first of the existing materials and then of deliberately fabricated heterostructures in which the known metallic delafossites are combined with the many insulating materials that have not been discussed in this review. There is, I believe, a realistic hope that such a programme could reveal a wealth of new physics and chemistry, and I hope that it is one of the medium- to long-term consequences of the fascinating foundational research that I have reviewed in this short article.

**Acknowledgements**

In the time since Yoshiteru Maeno and Hiroshi Takatsu first stimulated my interest in delafossite metals, I have benefited considerably from discussions on their physics with them and with a large number of other people, in particular Alexandra Gibbs, Maurits Haverkort, Deepa Kasinathan, Naoki Kikugawa, Phil King, Steve Kivelson, Philip Moll, Nabhanila Nandi, Helge Rosner and Chandra Varma. As well as productive discussions, my colleagues Pallavi Kushwaha, Seunghyun Khim and Marcus Schmidt contribute something even more important to my own group's efforts in the field, namely the high quality single crystals on which we perform our experiments. Finally, special thanks go to Clifford Hicks for some breakthrough measurements and important pieces of analysis, and to Veronika Sunko for many discussions as well as for her multiple careful readings of the draft, and for the host of perceptive and useful comments that she made on it.